\documentclass[pre,amsmath,amssymb,aps,reprint,groupedaddress,nofootinbib,raggedbottom,floatfix]{revtex4-2} 
\pdfoutput=1
\usepackage[T1]{fontenc}
\usepackage[utf8]{inputenc}
\usepackage{lmodern}
\usepackage{scalerel}

\usepackage[citecolor=blue, colorlinks=true, pdfusetitle, pdfauthor={Surajit Chakraborty, Vishnu V. Krishnan, Kabir Ramola, Smarajit Karmakar}, pdfsubject={Statistical Physics of Disordered systems}, pdfkeywords={Glass, Hessian, Eigenvalue}]{hyperref}

\begin{document}

\title{Enhanced Vibrational Stability in Glass Droplets}

\author{Surajit Chakraborty} 
\email{schakraborty@tifrh.res.in}

\author{Vishnu V. \surname{Krishnan}} 
\altaffiliation[Current address: ]{RCAST, University of Tokyo, 4-6-1 Komaba, Meguro, Tokyo 153-8904, Japan}
\email{vishnuvk@g.ecc.u-tokyo.ac.jp}

\author{Kabir Ramola} 
\email{kramola@tifrh.res.in}
\author{Smarajit Karmakar} 
\email{smarajit@tifrh.res.in}
\affiliation{Tata Institute of Fundamental Research, Hyderabad 500046, India}

\date{\today}

\begin{abstract}
We show through simulations of amorphous solids prepared in open boundary conditions that they possess significantly fewer low-frequency vibrational modes compared to their periodic boundary counterparts. Specifically, using measurements of the vibrational density of states, we find that the $D(\omega) \sim \omega^4$ law changes to $D(\omega) \sim \omega^\delta$ with $\delta \approx 5$ in two dimensions and $\delta \approx 4.5$ in three dimensions. Crucially, this enhanced stability is achieved when utilizing slow annealing protocols to generate solid configurations. We perform an anharmonic analysis of the minima corresponding to the lowest-frequency modes in such open-boundary systems and discuss their correlation with the density of states. A study of various system sizes further reveals that small systems display a higher degree of localization in vibrations. Lastly, we confine open-boundary solids in order to introduce macroscopic stresses in the system, which are absent in the unconfined system and find that the $D(\omega) \sim \omega^4$ behavior is recovered.
\end{abstract}


\maketitle

\section{Introduction}
In contrast to crystalline solids, whose vibrational states are well described by the Debye model, amorphous solids exhibit anomalous mechanical and thermal properties~\cite{zeller1971thermal, pohl2002low, argon1979plastic, falk1998dynamics, maloney2004universal, demkowicz2005liquidlike}. For a system in $d$-dimensions, the Debye model predicts a vibrational density of states (VDoS), $D(\omega) \sim \omega^{d - 1}$ arising from phonons~\cite{kittel2005introduction}. On the other hand, disordered and anharmonic systems possess an excess of low-frequency modes above the Debye prediction, termed the ``Boson peak''~\cite{inoue1991low, buchenau1984neutron, shintani2008universal, baggioli2019universal}. It has been suggested that the large specific heat of glasses and the plastic failure of amorphous solids are intimately related to these low frequency non-phononic vibrational modes~\cite{anderson1972anomalous, manning2011vibrational, chen2011measurement, widmer2008irreversible, xu2010anharmonic}. However,  a complete understanding of the structural and statistical properties of these vibrational modes has been elusive and remains an important current topic of interest in the field of disordered solids.

Several theoretical frameworks for the vibrations in disordered systems seek to model amorphous solids as an ensemble of anharmonic oscillators. Such a treatment is motivated by a disordered arrangement of particles containing local `soft spots' where the stiffness associated with a collective vibration is very small. The phenomenological ``Soft Potential Model'' (SPM) treats these regions as non-interacting oscillators and predicts a VDoS $D(\omega) \sim \omega^4$ for the lowest frequencies of stable inherent structures~\cite{il1987parameters, galperin1989localized, gurarie2003bosonic}. An extension to the SPM that includes effects of interactions between the anharmonic oscillators continues to retain the $\omega^4$ behavior of the VDoS~\cite{gurevich2003anharmonicity,gurevich2005pressure, parshin2007vibrational}. Other recent theoretical studies also predict a $D(\omega) \sim \omega^4$ at the lowest frequencies~\cite{ikeda2019universal,bouchbinder2021low,ji2019theory}. ``Fluctuating Elasticity Theory'' on the other hand predicts a low-frequency regime composed of extended modes that scale as $D(\omega) \sim \omega^{d+1}$ in $d$ dimensions~\cite{marruzzo2013heterogeneous, schirmacher2011some}. More recent work suggests non-affine displacements as the source of such non-Debye behavior~\cite{baggioli2022theory, szamel2022microscopic}. Certain mean field theories such as the ``Perceptron Model''~\cite{franz2015universal} and the ``Effective Medium Theory''~\cite{degiuli2014force} predict non-phononic vibrations with a $D(\omega) \sim \omega^2$ dependence. In this context, numerical investigations of amorphous solids in order to ascertain the nature of the low-lying excitations of such systems are of crucial importance.

Recent numerical studies of amorphous solids have identified a universal $D(\omega)\sim \omega^4$ scaling in the low-frequency regime of the VDoS across a broad class of simulated model systems~\cite{lerner2021low}. The universality of this non-phononic power-law at low frequencies has been established in $2, 3$ and $4$ dimensions~\cite{lerner2016statistics,richard2020universality,kapteijns2018universal}. In order to extract this behavior, studies have focused on small system sizes, where the lowest-frequency quasilocalized vibrations are well separated in energy from the first system-spanning phonon. This suggests that these quasilocalized modes (QLM) are primarily responsible for the power-law tail in the VDoS~\cite{kapteijns2018universal,paoluzzi2020probing}. Further, suppression of system-spanning vibrations using random pinning protocols has been shown to enhance the non-phononic spectra, thereby displaying a pronounced $D(\omega) \sim \omega^4$ behavior ~\cite{angelani2018probing}. Yet other studies have utilized measures such as participation ratios in order to isolate such quasilocalized modes that have been shown to contribute to the observed power-law behavior~\cite{mizuno2017continuum}. Simulations of various systems including ultrastable glasses~\cite{wang2019low}, silica models~\cite{bonfanti2020universal}, long-ranged models~\cite{das2020robustness}, finite-temperature systems~\cite{das2021universal} and random matrix models~\cite{stanifer2018simple}, all provide significant evidence of the ubiquity of the $\omega^4$ regime of the VDoS. However, other studies simulating amorphous solids report deviations from  the universal quartic law. High parent temperatures, poor annealing, and small system sizes have each been shown to result in power-law exponents that are less than $4$~\cite{lerner2017effect, lerner2020finite, lerner2022nonphononic}. Recent studies have further identified exponents of $3$ and $3.5$ in the low-frequency regime~\cite{wang2021low, wang2023scaling, wang2022density}. Confined three-dimensional thin films have also been shown to possess a low-frequency VDoS of $\omega^{3}$~\cite{yu2022omega}.

A crucial aspect often overlooked in simulations of amorphous solids is the residual shear stresses arising due to periodic boundary conditions (PBC)~\cite{dagois2012soft}. This implies that energy-minimized configurations of disordered systems under PBC are unstable to shear deformations. It has recently been shown that the low-frequency regime of the VDoS is modified to $D(\omega) \sim \omega^5$ when considering ensembles stable to simple-shear perturbation~\cite{krishnan2022universal}. Significantly, such an increase in the exponent points to a correlation between the shear stability of the system and a reduction in the propensity for localized vibrations. Given these observations, a natural question that arises pertains to the consequences of stabilization against all possible deformations. Solids formed under \emph{open boundary conditions} (OBC) are a suitable candidate since all elements of their pressure tensor are identically zero for each configuration. Such a state is permitted by a lack of confinement at the boundaries. It is important to note that the OBC system is no longer isotropic and introduces possible radial heterogeneity in structure as well as relaxation dynamics. Furthermore, unlike systems under PBC, phonons in OBC are not required to obey the artificial symmetries enforced by the periodicity. Consequently, solids under OBC can accurately capture features of natural solids, including surface as well as system-size effects. 
\begin{figure*}[t!]
\includegraphics[width=0.95\textwidth]{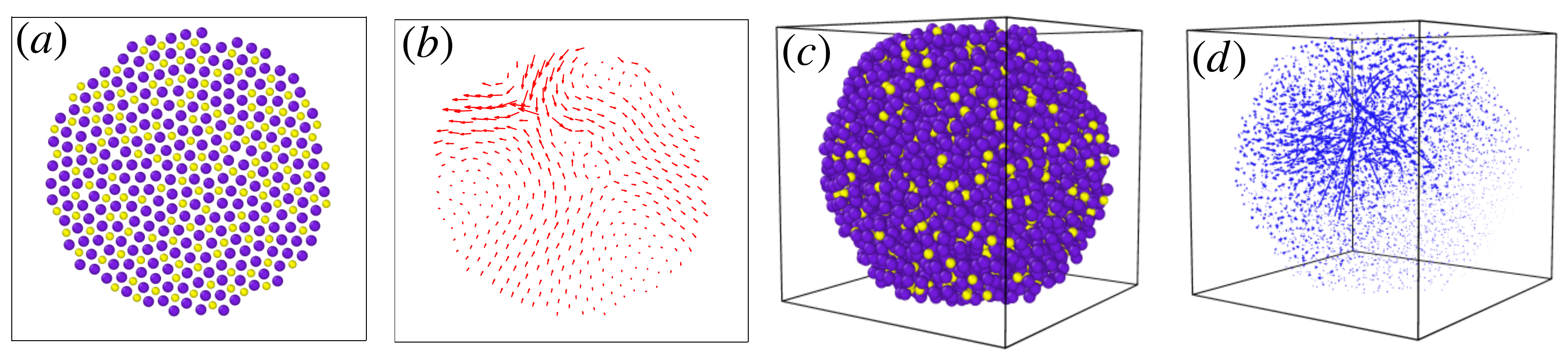}
\caption{
\textbf{(a)} A two-dimensional solid generated from a $400$ particle, circular `cut-out' of a liquid configuration, via Damped Dynamics minimization under open boundary conditions. \textbf{(b)} The first non-zero vibrational mode of the two-dimensional solid with a participation ratio of approximately $0.26$ and a frequency $\omega \approx 4.34 \times 10^{-1}$, displaying typical quasilocalized characteristics. \textbf{(c)} A three-dimensional solid configuration of $4096$ particles under open boundary conditions generated using Damped Dynamics energy minimization. \textbf{(d)} The first non-zero vibrational mode of the three-dimensional solid with a participation ratio of $0.0068$ and a frequency $\omega \approx 4.04 \times 10^{-1}$. Note that the solid borders of the figures are to aid depth perception, especially in three dimensions. The systems themselves have no boundaries.}\label{fig_mode}
\end{figure*}
In this paper, we report a characterization of the VDoS of open boundary amorphous solids. The low-frequency vibrational properties of open systems remain relatively unexplored~\cite{tanguy2002continuum}, and our study forms the first such computational examination of the localized modes of open-boundary amorphous solids. We observe the low-frequency vibrational spectrum of a simulated model amorphous solid under open boundary conditions to be of the form $D(\omega) \sim \omega^\delta$ with $\delta > 4$, both in two and three dimensions. Since an increase in the exponent implies a reduction in the degree of vibrational localization, the corresponding solid ensemble may be said to possess enhanced stability. Notably, the model system displays such a VDoS only when configurations are \emph{annealed} to their inherent structures. On the other hand, commonly used quenching protocols lead to ensembles with $\delta \sim 4$. Interestingly, at large enough system sizes, the exponent saturates to a value $\delta = 5$ in 2D and $\delta = 4.5$ in 3D. A detailed investigation of the average stress profile of the solids allows us to identify a surface layer that suggests a source of the system-size effects. Lastly, we also observe that confining open boundary solids under a harmonic trap recovers the $D(\omega) \sim \omega^{4}$ behavior of PBC systems, confirming the role of stresses in the stability of solids.

The outline of the paper is as follows. Section \ref{sec_model} describes the numerical protocol we use to generate stable open-boundary solids. In Section \ref{sec_stability}, we demonstrate the effect of different minimization protocols on the generation of stable solids. In Section \ref{sec_system_size}, we present data to demonstrate the effect of the system size on the vibrational spectrum. In Section \ref{sec_bulk_sur}, we provide an analysis of the source of the stabilization. In Section \ref{sec_confinement}, we examine the effect of confinement on the vibrational spectrum of solids. Finally, we conclude and provide directions for future research in Section \ref{sec_conclusion}.


\section{Model and Simulation Details}\label{sec_model}
In order to simulate the behavior of amorphous solids in open boundary conditions, we use models with attractive interactions between the particles. Specifically, we use variants of the canonical Kob-Andersen Lennard-Jones model~\cite{kob1995testing,bruning2008glass} in two and three dimensions. The model consists of binary mixtures of particles in number ratios $65:35$ in two dimensions (2D) and $80:20$ in three dimensions (3D), with an interaction potential:
\begin{footnotesize}
\begin{equation}
    V_{\alpha \beta}(r) = 4 \epsilon_{\alpha \beta}\left[\left(\frac{\sigma_{\alpha \beta}}{r}\right)^{12}-\left(\frac{\sigma_{\alpha \beta}}{r}\right)^{6}+\sum_{i=0}^{2} c_{2 i}\left(\frac{r}{\sigma_{\alpha \beta}}\right)^{2 i}\right]
\end{equation}
\end{footnotesize}
where $\alpha, \beta \in\{A, B\}$ correspond to the two types of particles resulting in three types of interactions. The potential is cut off at a distance $r^{\alpha\beta}_c=2.5 \sigma_{\alpha\beta}$. $\sigma_{AA}=1$ is the unit of length, and $\epsilon_{AA}=1$ is the unit of energy (Boltzmann's constant being unity). The remaining parameters are $\epsilon_{AB}=1.5$, $\epsilon_{BB}=0.5$, $\sigma_{AB}=0.8$, and $\sigma_{AB}=0.88$ with all masses set to unity $m=1.0$. We equilibrate liquid configurations at high temperatures ($T=0.55$ in both two and three dimensions) by performing constant temperature molecular dynamics simulations under periodic boundary conditions. The liquid is then cooled to a low temperature ($T=0.2$) at a rate $\dot{T} = 10^{-4}$. In order to achieve a density close to that of a system under open boundary conditions, we perform zero-pressure molecular dynamics simulations using the isobaric ensemble.

Finally, in order to study solid droplets under OBC, we begin by cutting out a circular (spherical in 3D) region of the zero-pressure liquid. We ensure that all the droplet samples contain a fixed number of particles, say $N$, by selecting the $N$ closest particles to the center of mass of the corresponding liquid sample referred to henceforth as a ``cut-out'' sample. Solids are configurations that resist deformations, which correspond to the inherent structures of the amorphous configuration of particles. We generate such structures by performing an energy minimization of the particle-position degrees of freedom using three different protocols, namely: (i) Molecular dynamics simulations in the presence of dissipative viscous drag in the athermal limit (Damped Dynamics) ~\cite{mandal2020extreme}, (ii) FIRE~\cite{bitzek2006structural} and (iii) the non-linear conjugate gradient algorithm~\cite{shewchuk1994introduction}. Note that we do not equilibrate the cut-out sample but instead directly quench or anneal through the energy minimization protocol.

Simulations of the largest system-size three-dimensional zero-pressure PBC liquids are performed using LAMMPS~\cite{thompson2022lammps}. The vibrational properties of the solid configurations are probed through the Hessian matrix as defined in Eq.~\eqref{eqn_hess}. We perform eigenvalue computations using the Intel Math Kernel Library~\cite{intelmkl} sparse solver routine \texttt{mkl\_sparse\_d\_ev}. Below, we discuss the three protocols we employ in order to find the minimum-energy amorphous configurations.

\subsection{Damped Dynamics}
In the Damped Dynamics (DD) protocol, a  dissipative viscous drag in the absence of a temperature bath or other energy input serves as a method to find the stationary state of the conservative force fields. Such an evolution of the system is described by the equation of motion 
\begin{equation}
  \frac{d\vec{v}_i}{dt}=\vec{F_i}-\gamma \vec{v}_i  
\end{equation}
where $\vec{F}_i$ is the conservative force on particle $i$  due to interactions with other particles, $\vec{v}_i$ is the instantaneous velocity of the particle $i$ and $\gamma$ is the viscous damping parameter. A stable steady-state solution of these equations corresponds to a minimum energy configuration of the system of particles with $\vec{F}_i = 0$ and $\vec{v}_i = 0$ for all particles. We employ a velocity Verlet integration scheme modified to incorporate velocity-dependent accelerations. These damped equations of motion allow the system to traverse the basins of multiple inherent states before selecting and settling in a more stable minimum. 
We employ small damping constants so that the system is able to execute such a relaxation. At large values of $\gamma$, the dynamics would be equivalent to a steepest-descent minimization of the energy. Unless explicitly mentioned the value of the damping constant $\gamma$ is $0.1$. 

\subsection{FIRE Minimization}
The Fast Inertial Relaxation Engine (FIRE) is an efficient protocol designed to find local minima of multidimensional functions~\cite{bitzek2006structural}. The procedure involves numerically integrating a dynamical equation with variable viscous damping and, additionally, a gradient director. The equation of motion for each particle is
\begin{equation}
    \frac{d \vec{v}}{d t}=\vec{F}(t)-\gamma(t)|v(t)|[\hat{v}(t)-\hat{F}(t)] \label{eqn_fire}
\end{equation}
where $\hat{F}(t)$ refers to unit vector along $\vec{F}(t)$. These two aspects of the FIRE algorithm that lend it its speed also prevent the system from leaving its original energy basin. Therefore the inherent structure obtained is primarily a function of the equilibrium sampling and is not dependent on the stability of the minimum of the basin.

\subsection{Conjugate Gradient Minimization}
We also test the stability of configurations obtained via a conjugate gradient (CG) minimization scheme ~\cite{shewchuk1994introduction}. We implement the non-linear conjugate gradient for the open boundary system using the Polak-Ribiere method to obtain the updated direction and the secant minimization method in order to determine the step size of the line search. As in the case of the FIRE algorithm, the conjugate gradient leads the system to its nearest minimum.

\begin{figure*}[t!]
\begin{center}
\includegraphics[width=0.95\textwidth]{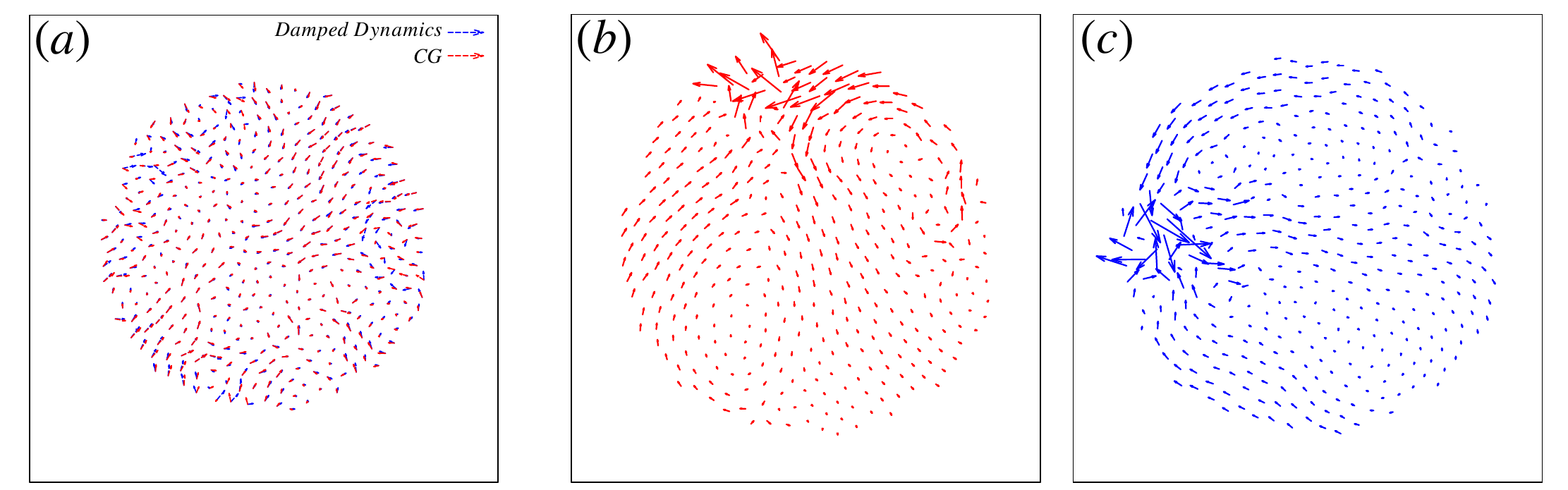}
\caption{\textbf{(a)} Displacements incurred by each particle from the same liquid configuration when undergoing energy minimization through two different protocols, namely Conjugate Gradient and Damped Dynamics. The eigenvectors corresponding to the lowest frequency vibrational modes of the solid configuration obtained via \textbf{(b)} Conjugate Gradient and  \textbf{(c)} Damped Dynamics minimization protocols. Under a quartic Taylor expansion of the landscape along the modes, the energy minimized configuration achieved via \textbf{(b)} CG is unstable and by \textbf{(c)} DD is stable.\label{fig_minimization}}
\end{center}
\end{figure*}

\section{Stable Vibrational Modes}\label{sec_stability}

Harmonic vibrations of a solid can be probed by diagonalizing the ``Hessian Matrix'', defined as 
\begin{equation}
    \left.\mathbf{H}_{i j} \equiv \frac{\partial^{2} U\left(\mathbf{r}_{1}, \mathbf{r}_{2}, \ldots \mathbf{r}_{n}\right)}{\partial \mathbf{r}_{i} \partial \mathbf{r}_{j}}\right|_{\left\{\mathbf{r}_{i}^{0}\right\}}, \label{eqn_hess}
\end{equation}
where $U\left(\mathbf{r}_{1}, \mathbf{r}_{2}, \ldots \mathbf{r}_{n}\right)$ is the total potential energy of the system and $\mathbf{r}_{i}$ denotes the position of particle $i$. The eigenvectors $(\psi_k)$ of the Hessian matrix correspond to the cooperative displacements of the particles participating in harmonic vibrations (termed normal modes), and the corresponding eigenvalues $(\lambda_k)$ represent the frequency of that vibration. The system, when perturbed along a normal mode, will perform a pure oscillation with a frequency $\omega_k = \sqrt{\lambda_k}$. The Hessian matrix evaluated at an energy minimum of the landscape is positive semi-definite with zero modes corresponding to the global invariants of the Hamiltonian (Goldstone modes). For example, a 3D system under PBC possesses three translational degrees of freedom, and the corresponding Hessian, therefore, has three zero-modes. Systems under OBC, on the other hand additionally also possess rotational degrees of freedom. In general, in $d$ dimensions, such a solid has $\frac{d(d+1)}{2}$ zero modes corresponding to $d$ translational modes and $\frac{d(d-1)}{2}$ rotational modes. The vibrational density of states (VDoS) of a system of ${N}$ particles in $d$ dimensions is defined as 
\begin{equation}
    D(\omega)=\frac{1}{N d} \sum_{i=1}^{Nd} \delta\left(\omega-\omega_{i}\right)
\end{equation}

In Fig.~\ref{fig_mode}~\textbf{(a)} and~\textbf{(b)} we display a typical open-boundary 2D solid configuration of $N = 400$ particles and its first nonzero mode corresponding to a frequency $\omega = 3.24107\times10^{-1}$ with a participation ratio (defined in Eq.~\eqref{eqn_PR}) of $1.313714\times10^{-1}$. Similarly, in Fig.~\ref{fig_mode}~\textbf{(c)} and~\textbf{(d)} we show a typical 3D solid of $N = 4096$ particles and its first mode corresponding to a frequency $\omega=4.039514\times10^{-1}$ and participation ratio of $6.792794\times10^{-3} $. These low-frequency modes are quasilocalized with a small number of particles displaying large participation by forming the core of the vibration. Interestingly, these eigenmodes resemble the quadrupolar modes observed in the low-frequency regime of the vibrational density of states of structural glass formers under PBC~\cite{lerner2021low}.

\subsection{Anharmonic Stability Analysis}\label{sec_anhar_stability}
While the Hessian characterizes the curvature of the energy minimum and, thereby, the frequency of vibrations, an anharmonic analysis is necessary in order to understand the vicinity of the energy minimum better. The energy near the minimum may be approximated up to the fourth order as
\begin{equation}
    \delta U(s)=\frac{1}{2!} B_2 s^2+\frac{1}{3!} B_3 s^3+\frac{1}{4!} B_4 s^4.\label{eqn_quartic_expan}
\end{equation}
where $s$ is the scalar distance of particles from the minimum along a given direction in the energy landscape. If the particles are displaced along an eigenvector ($\psi$) of the Hessian ($H$),
\begin{equation}
    B_2=\psi \cdot H \cdot \psi    
\end{equation}
is the eigenvalue of the Hessian, and 
\begin{equation}
  B_3=\psi_i\left[\frac{\partial^3 U}{\partial r_i \partial r_j \partial r_l}\right] \psi_l \psi_j, 
\end{equation}
\begin{equation}
    B_4=\psi_i \psi_l\left[\frac{\partial^4 U}{\partial r_i \partial r_l \partial r_k \partial r_j}\right] \psi_k \psi_j 
\end{equation}
quantify the non-linearity of the energy landscape along that mode.

Given a vibration in the energy landscape as defined by an eigenvector, the corresponding energy minimum is said to be `stable' provided that the minimum is the deepest in its neighborhood. Under the quartic approximation of the energy near the minimum, such stability is achieved when~\cite{gurarie2003bosonic}
\begin{equation}
    B_{3}^{2} \leq 3 B_{2} B_{4}.
    \label{eqn_stability_criterion}
\end{equation}

\subsection{Effect of Minimization Protocol}
Simulated models studying the properties of amorphous solids have typically been systems under periodic boundaries. It has been shown that in such systems, the CG minimization protocol generates configurations with stable minima, provided that configurations are obtained from a sufficiently low parent temperature~\cite{lerner2021low}. 
However, the open boundary configurations obtained by quenching the cut-outs of zero-pressure PBC systems contain, among them, some inherent structures at unstable minima. We, therefore, explore an energy-minimization protocol that anneals the system into an ensemble comprised of primarily stable minima.

\begin{figure}[t!]
\begin{center}
\includegraphics[width=0.48\textwidth]{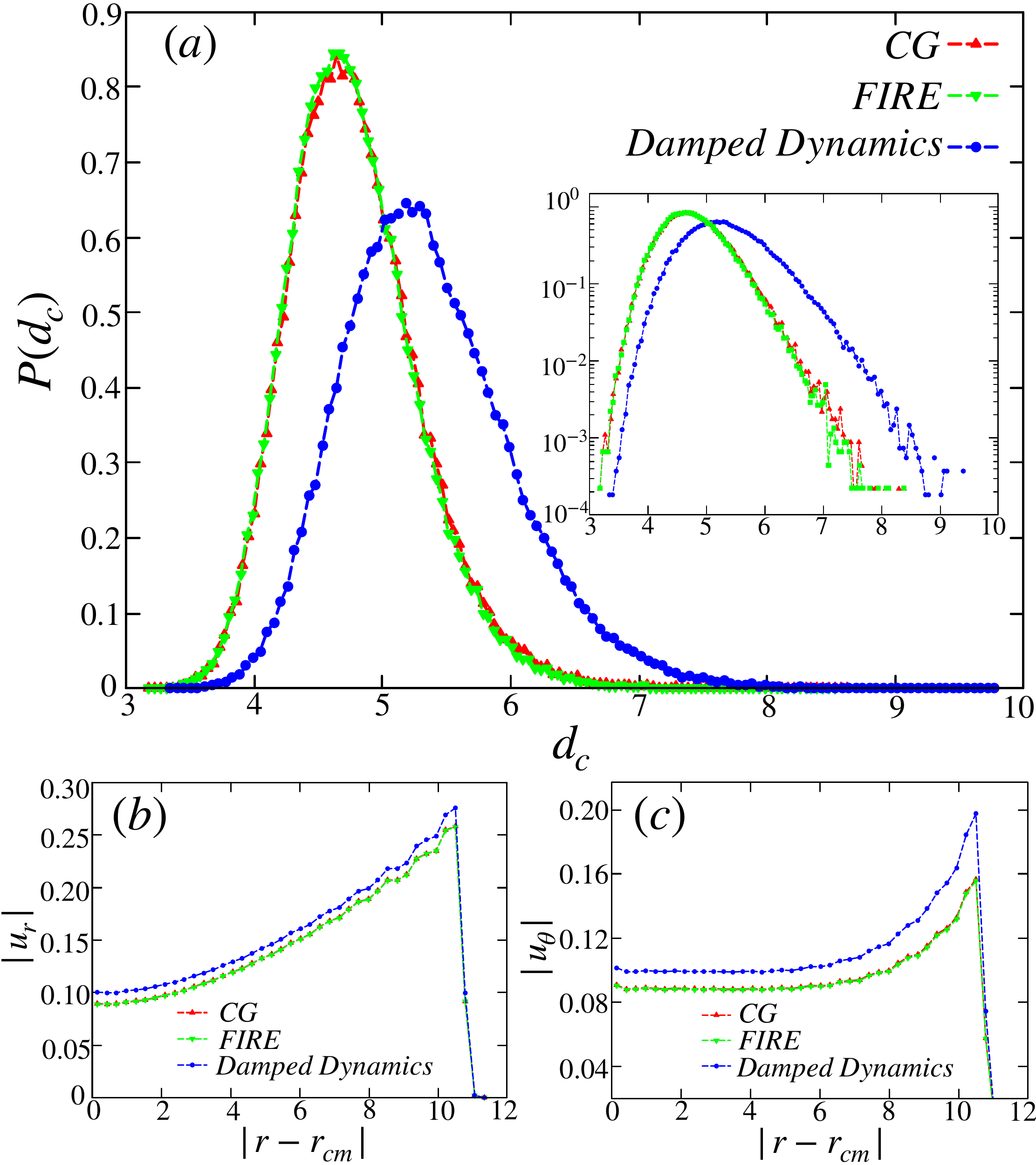}
\caption{\textbf{(a)} Distribution of the distances between the open boundary liquid and solid configurations for different minimization protocols. Particles in configurations that underwent Damped Dynamics minimization show relatively larger displacements. \textbf{(Inset)} The distribution plotted in log-scale to highlight the difference at large $d_{c}$. Average \textbf{(b)} radial and \textbf{(c)} tangential displacements incurred by particles at different radial distances. Damped Dynamics produces a large displacement of the particles in the tangential direction, further enhanced near the surface of the liquid as compared to other minimization protocols.}\label{fig_dispacemnent_min_proto}
\end{center}
\end{figure}

We find that different energy minimization protocols drive the system to different minima. In Fig.~\ref{fig_minimization}~(a), we show the displacements incurred by each particle from a particular liquid configuration when undergoing energy minimization through two different protocols, namely Damped Dynamics and Conjugate Gradient. While both procedures yield inherent structure configurations, they do not correspond to the \textit{same} minimum. Most significantly, Damped Dynamics finds minima in a relatively expanded region in the energy landscape as compared to CG and FIRE. In Fig.~\ref{fig_dispacemnent_min_proto} (a) we present corroborating numerical evidence with particles undergoing relatively larger displacements in the case of Damped Dynamics. In order to qualify this difference, we compute the distance between the cut-out liquid configuration and the energy-minimized solid configuration as 
\begin{equation}
 d_c=\sqrt{\sum_{i=1}^N|\vec{r}_i^L-\vec{r}_i^S|^2 }
\end{equation}
where $\vec{r}_i^L$ and $\vec{r}_i^S$ represent the position of particle $i$ in the liquid and solid configurations respectively. In terms of the radial and tangential displacement incurred by particles at different radial distances,
\begin{equation}
 d_c^2=\int_{r=0}^{R}r dr (|u_{r}(r)|^2+|u_{\theta}(r)|^2)
\end{equation}
where $u_{r}(r)$ and $u_{\theta}(r)$ are the radial and tangential displacement of particles at a radial distance $r$ from the center of mass of the liquid droplet. In Fig.~\ref{fig_dispacemnent_min_proto} we show the average \textbf{(b)} radial and \textbf{(c)} tangential displacement at different radial distances of the liquid configurations. For the case of Damped Dynamics minimization, particles near the surface undergo comparatively larger displacement in the tangential direction, which results in a stable inherent structure. Since the viscosity of the surrounding medium determines the exact extent of this search space, we use an appropriately chosen damping, optimizing for both improved search area as well as speed.

We now consider the vibrational properties of the minima obtained through the various energy minimization protocols. In Fig.~\ref{fig_minimization}, we show the lowest frequency vibrational mode of a solid configuration obtained via \textbf{(b)} CG and \textbf{(c)} Damped Dynamics minimization of the same liquid configuration. Although both modes show signatures of similar quadrupolar vibrations, the inherent structure corresponding to \textbf{(a)} does not satisfy the stability criterion defined in Eq.~\ref{eqn_stability_criterion}.

\begin{figure}[t]
\begin{center}
\includegraphics[width=0.48\textwidth]{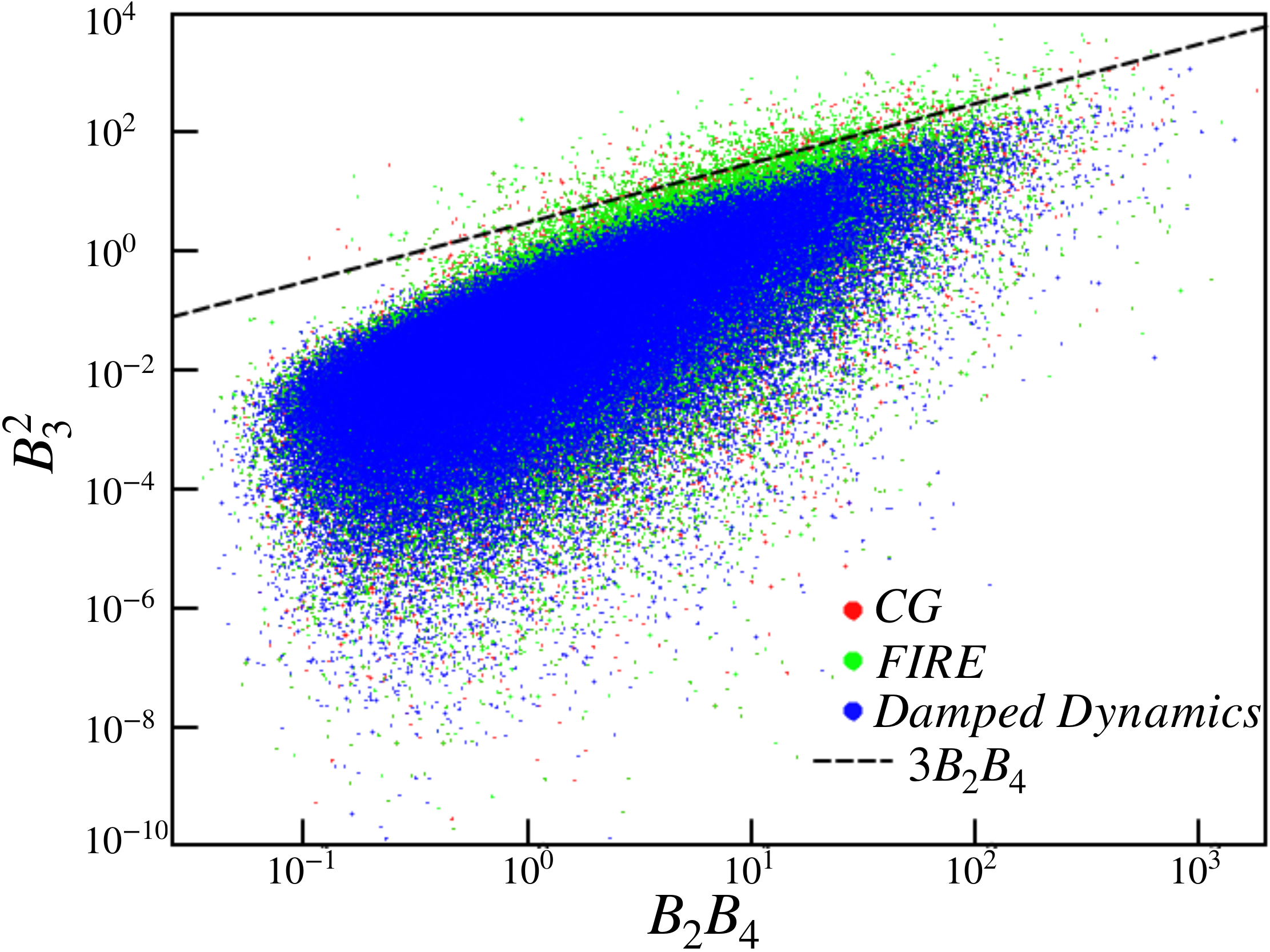}
\caption{Scatter plot of $B_3^2$ against $B_2B_4$ with a sample size of $2\times10^5$ solid configurations. The dashed line corresponds to $B_3^2 = 3B_2B_4$ with the region of stability lying below. CG and FIRE minimization protocols produce some inherent structures at unstable minima, whereas Damped Dynamics primarily generate stable minima.}\label{fig_scatter}
\end{center}
\end{figure}

An important difference between the ensembles of minima obtained via the various protocols lies in their vibrational stability as defined in Eq.~\ref{eqn_stability_criterion}.
Through Fig.~\ref{fig_scatter}, we assess the stability of each of the open-boundary solid configurations across the different minimization protocols. We use the quartic Taylor expansion as illustrated in Eq.~\ref{eqn_quartic_expan} along the direction specified by the lowest frequency mode in order to extract the stability of the minimum. Interestingly, the CG and FIRE minimization protocols present many unstable inherent structures indicated by the data points above the stability line corresponding to $(B_3^2 = 3B_2B_4)$. At the same time, the Damped Dynamics minimizer produces predominantly stable inherent structures as shown by the data points being well within the regime of stability. This suggests that the Damped Dynamics minimizer allows the system to find a locally deeper minimum.

\begin{figure}[t]
\begin{center}
\includegraphics[width=0.48\textwidth]{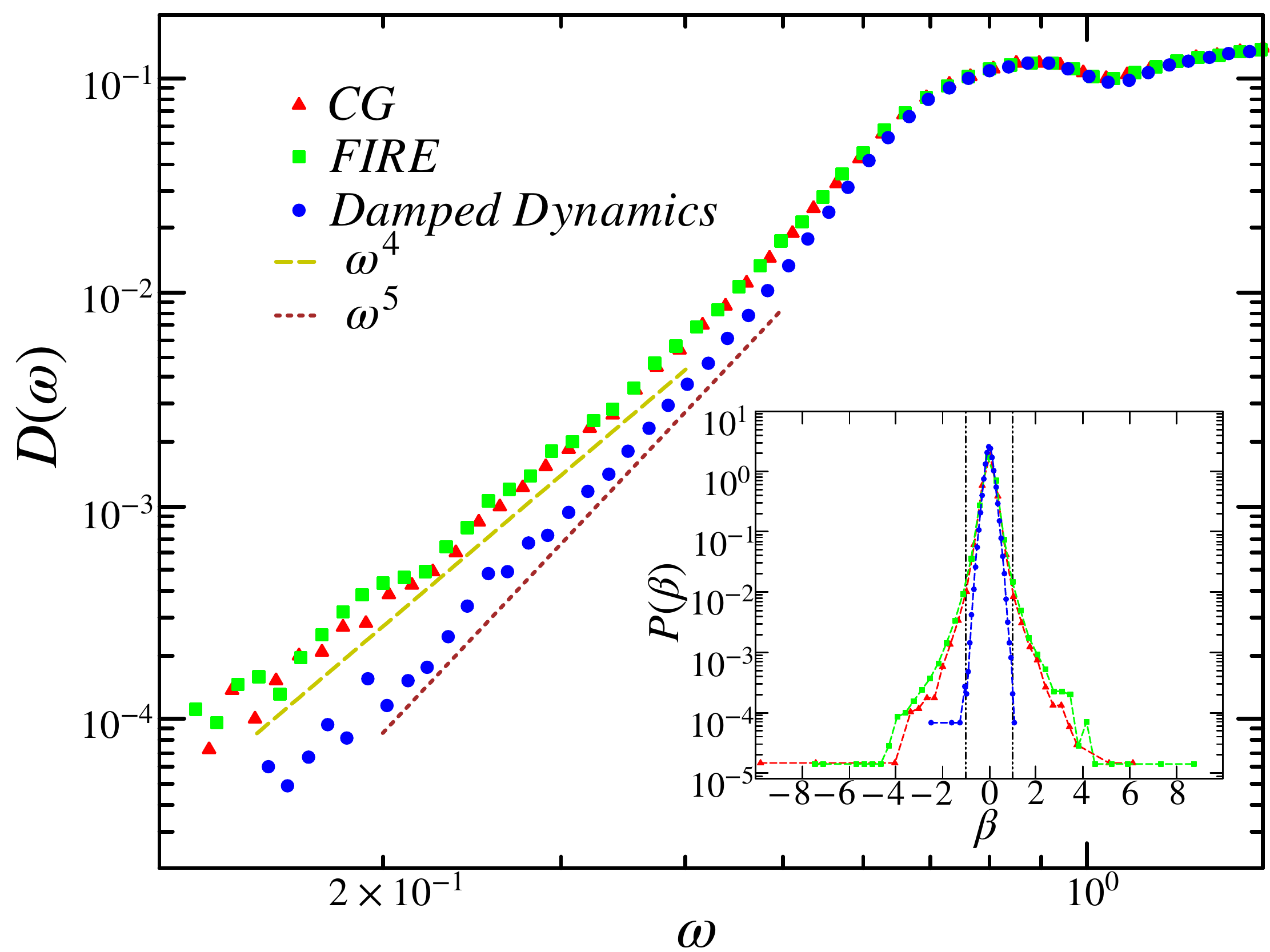}
\caption{Distribution of the low-frequency vibrational modes of a $400$-particle system in two dimensions under open boundary conditions. The histogram is generated by sampling the first $100$ non-zero eigenvalues of $3\times10^5$ configurations. The solid ensembles generated using CG and FIRE minimizers display a low-frequency behavior of $\omega^{4}$, whereas using the Damped Dynamics minimizer displays $\omega^{5}$. (\textbf{Inset}) Distribution of the stability factor $\beta = B_3/\sqrt{3B_2B_4}$ for solids obtained via the various minimization protocols. The dashed vertical lines correspond to $\left| \beta \right| = 1$, with the enclosed regions containing stable configurations.}\label{fig_stability_dos}
\end{center}
\end{figure}

Since the choice of minimization protocol affects the stability of minima, we next examine its effects on the vibrational spectrum. In Fig.~\ref{fig_stability_dos}, we plot the distribution of the low-lying vibrational frequencies of two-dimensional solid ensembles. While the VDoS of CG and FIRE minimized ensembles display the erstwhile universal low-frequency power law $D(\omega) \sim \omega^4$~\cite{lerner2016statistics, lerner2021low}, ensembles generated via the Damped Dynamics minimizer show a power law of $\omega^5$ in the low-frequency regime, as has been seen previously in shear-stabilized systems~\cite{krishnan2022universal}. Generically the smaller the lowest frequency mode, the closer the system is to instability. An increase in the power of the VDoS from $4$ to $5$ corresponds to a reduction in the proportion of such small-valued modes. Therefore, such a change is a signature of a transformation to a more stable solid ensemble enabled by an appropriate choice of an annealing energy minimizer.

In the inset of Fig.~\ref{fig_stability_dos}, we plot the distribution of
\begin{equation}
    \beta = \frac{B_3}{\sqrt{3B_2B_4}},
\end{equation}
where states within the dashed lines $(|\beta|<1)$ correspond to stable minima. We find that the stability condition is best satisfied by configurations obtained via Damped Dynamics minimization. Thus, we observe that the model system, under open boundary conditions, displays a correlation between stable minima and the vibrational stabilization of the corresponding solid ensembles.

\section{System Size Effects}\label{sec_system_size}  
It is well-known that the low-frequency vibrational spectrum of amorphous solids contains system-spanning phonons in addition to localized vibrations~\cite{lerner2021low}. The long-wavelength phonons in a solid with linear dimension $L$, under periodic boundary conditions, possess a frequency that varies with system size as $L^{-1}$. This decrease in the frequency of the phonons with increasing system size leads to difficulty in the characterization of quasilocalization at low frequencies. At the same time, the regime of the VDoS that is most amenable to the study of quasilocalized behavior are the frequencies below the first phonon~\cite{gartner2016nonlinear}. As the phonon frequencies become comparable to the frequencies of quasilocalized modes at larger system sizes, it is then necessary to perform a disorder averaging over an increasingly large number of samples in order to distinguish these modes. This becomes computationally infeasible at very large system sizes. Additionally, QLMs also display finite size effects when the extent of these modes becomes comparable to the size of the system~\cite{lerner2020finite}. The softness of QLMs, therefore, becomes independent of system size once it surpasses the length scale of the localized vibration. In this context, very small system sizes have been shown to possess an excess of quasilocalized vibrations with $D(\omega) \sim \omega^{\delta}$ where $\delta < 4$~\cite{lerner2017effect,lerner2020finite,lerner2022nonphononic}, as the finite size effects dominate in such situations.

Solids prepared under open boundary conditions can display additional surface effects in their low-frequency vibrational spectrum, as boundary effects not present in PBC systems also play a role. The stiffness of particle motion near the surface of the solid is small compared to the bulk of the material. Therefore the eigenmodes are softer near the boundary of the solid. It is thus natural to expect that in small open boundary solids, there will be stronger surface effects than in larger solids. This occurs in addition to the finite-size effects arising from the interplay between the size of the system and the size of the quasilocalized modes. It is therefore important to study the variation of the VDoS with system size under open boundary conditions in order to identify regimes of validity of the enhanced stability.

A quantitative characterization of the surface effects may be performed through a measurement of the stress distribution within such solids. Moreover, since the low-lying VDoS is crucially sensitive to the macroscopic stress in the amorphous structure~\cite{krishnan2022universal}, it is reasonable to assume that the finite size effects displayed by the stress profiles also carry over to the VDoS. In PBC systems, the average stress displays homogeneity over the sample, and therefore the local elastic behavior is also independent of the location in the solid. However, as the macroscopic stress is precisely zero in open boundary solids, any bulk stress is necessarily counterbalanced by the stresses on a ``boundary layer''. Such a distinct surface region possessing a typical thickness is yet another source of finite-size effects. Additionally, the local elasticity properties of the material are correlated with the local stress distribution, with particles near the boundary displaying lower stiffness in their interaction. In this context, we study the boundary stress layer in solids prepared under open boundary conditions, in order to systematically analyze the finite size effects present in their low-frequency VDoS.

\subsection{Stress Distribution and the Boundary-layer}

\begin{figure}[t!]
\begin{center}
\includegraphics[width=0.5\textwidth]{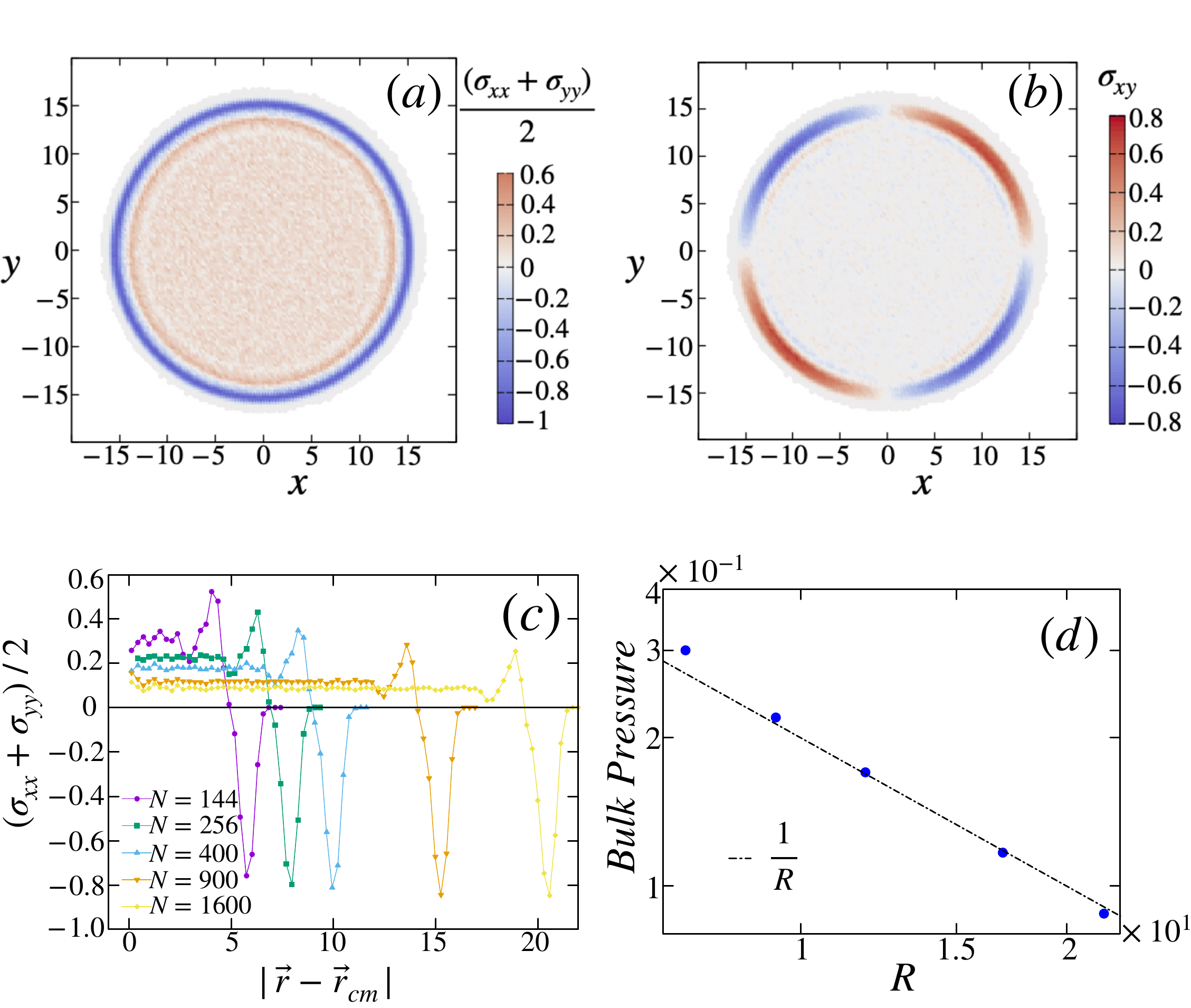}
\caption{Spatial distribution of the \textbf{(a)} pressure and \textbf{(b)} shear stress in solids of system size $N=900$. The stress profile is constructed by coarse-graining over bins of dimension $0.28~\sigma_{AA}\times0.28~\sigma_{AA}$ and averaging over $10^5$ solid configurations. \textbf{(c)} Radial distribution of pressure at different system sizes. Boundary layers approximately five particles thick are observed in all systems, independent of size. \textbf{(d)} Bulk pressure as a function of system radius. We observe a $\frac{1}{R}$ scaling of the bulk pressure where $R$ is the radius of the solid droplet.
}\label{fig_stress_profile}
\end{center}
\end{figure}
In Fig.~\ref{fig_stress_profile} we describe the stress distributions which highlight the existence of a surface layer in an open boundary solid. Specifically, in Fig.~\ref{fig_stress_profile}~\textbf{(a)} and~\textbf{(b)}, we plot the spatial distributions of the pressure and shear stress respectively, as seen in open boundary solids comprised of $900$ particles. These stress profiles are constructed by averaging over $10^5$ amorphous configurations and using coarse-graining boxes each of dimension $0.28~\sigma_{AA} \times 0.28~\sigma_{AA}$. The typical radius of a configuration is approximately $15~\sigma_{AA}$ resulting in about $9000$ bins. Fig.~\ref{fig_stress_profile}~\textbf{(c)} plots the radial distribution of the pressure for amorphous solids of various sizes. The stress distributions display clear indications of a boundary layer, with the effects of the surface spanning over approximately $5~\sigma_{AA}$. We also find that while the pressure in bulk is sensitive to the total size of the droplet, the thickness of the boundary layer is largely independent of the system size.

The microscopic stress field ($\sigma_{ij}(\mathbf{r}))$ can be expressed as \begin{equation} \sigma_{i j}(\mathbf{r})=\sigma_{i j}^0(\mathbf{r})+\delta \sigma_{i j}(\mathbf{r}), \label{eqn1_appen_3} \end{equation} where $\sigma_{i j}^0(\mathbf{r})$ represents the stress tensor of the initial liquid configuration and $\delta \sigma_{i j}(\mathbf{r})$ is the change in the stress tensor which results from the process of energy minimization. The original distribution of stress $\sigma_{ij}^0$ in the liquid is dependent on the preparation protocol. The effect of removing a circular cut-out from the zero-pressure liquid configuration gives rise to a non-trivial stress distribution near the boundaries of the cut-out. This excess stress causes the particles near the edges to be displaced further inwards (due to the attractive nature of the interaction) near the boundaries. This leads to a new stress-balanced state, \begin{equation}   \partial_{i} {\sigma_{i j}(\mathbf{r})}=0,\label{eqn1_appen_2} \end{equation} with $\delta\sigma_{ij}$ displaying larger changes near the boundary. The condition of mechanical equilibrium at the boundary of the solids is satisfied when the normal force acting on the surface due to the internal pressure exactly balances the surface tension force. Our numerical observations suggest that the stress profile within the surface layer is largely independent of the system size, leading to a constant surface tension across different system sizes. This suggests the bulk pressure will decrease as the radius of the solid is increased. In Fig.~\ref{fig_stress_profile}~\textbf{(d)}, we plot the bulk pressure for different system-size solids. We observe a $P \sim \frac{1}{R}$ scaling of the bulk pressure with the radius of the solid droplet $R$, which is consistent with a ``Laplace Law'' for the pressure in a droplet~\cite{behroozi2022fresh}.
 
\subsection{Vibrational Density of States}
Having characterized the boundary stress layer, we next turn to the system size dependence of the vibrational density of states. We expect the effects of the boundary to be significant only in system sizes for which the width of the boundary is comparable to the radius of the droplet. In order to discern the lengthscales corresponding to such a crossover, we analyze solids of various system sizes with particle numbers ranging from $N = 144$ to $1600$ in two dimensions and from $N = 216$ to $10000$ in three dimensions. 

\begin{figure}[t!]
\begin{center}
\includegraphics[width=0.48\textwidth]{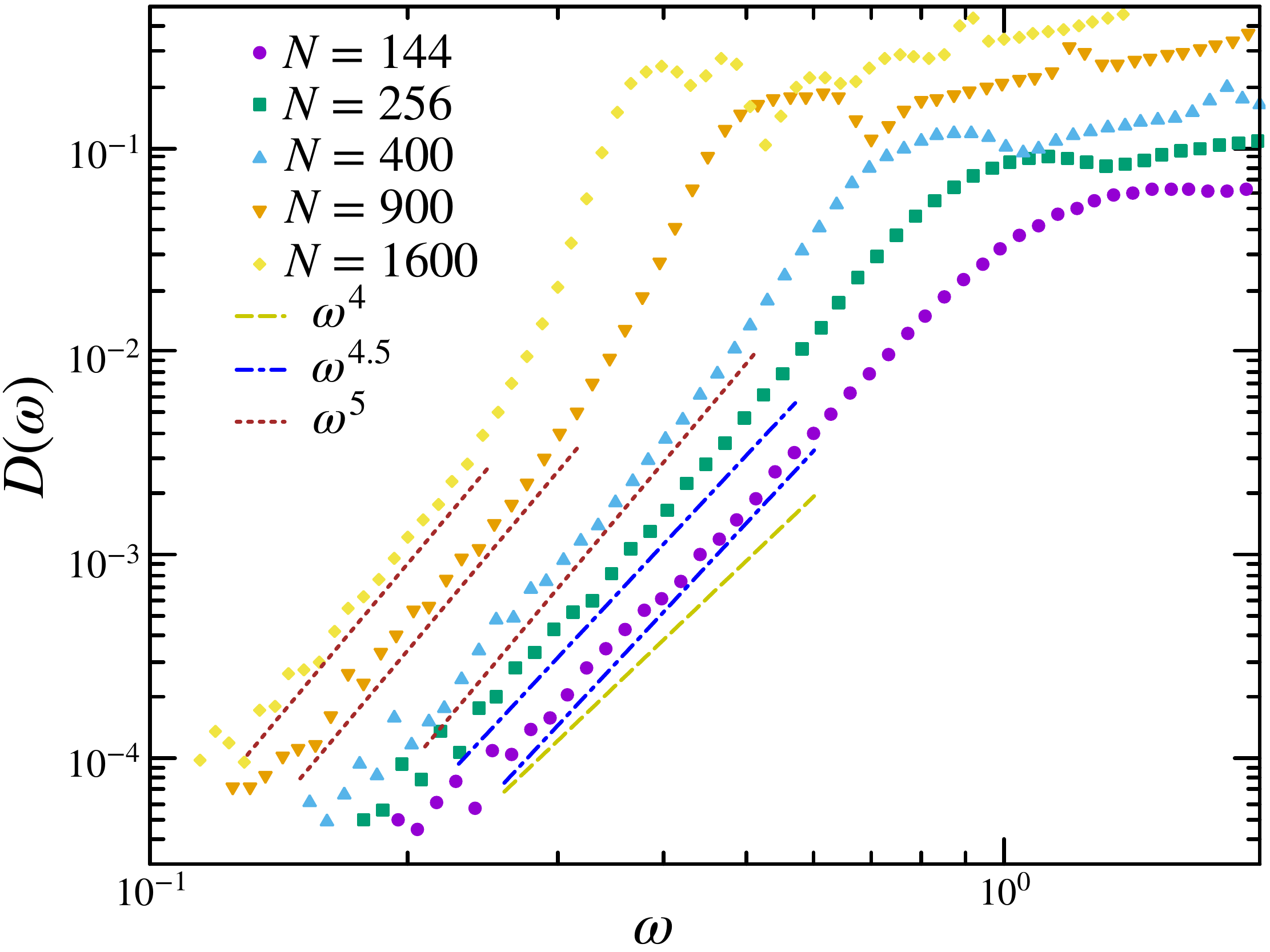}
\caption{System size dependence of the low-frequency vibrational spectrum of open boundary solids in two dimensions. We have sampled the first $100$ non-zero eigenvalues of $2\times10^5$ configurations for system sizes $N = 144$ and $256$. The histograms for the larger system sizes are constructed using $5\times10^5$ configurations. At small system sizes, the surface effects are more pronounced leading to more soft modes in their vibrational spectrum. They display a $D(\omega) \sim \omega^{\delta}$ with $\delta<5$ in the low-frequency regime. Such effects are reduced upon increasing the size with large systems displaying a $D(\omega)\sim\omega^5$.}\label{fig_system_size}
\end{center}
\end{figure}

In Fig.~\ref{fig_system_size}, we report the effect of system size on the low-frequency vibrational spectrum of open boundary solids in two dimensions. The small systems $(N = 144, 256)$ display a higher degree of softness in their vibration, as can be seen from their low-frequency behavior of $D(\omega) \sim \omega^{4.5}$. The VDoS of systems with particle numbers equal to and larger than $400$ appear to exhibit a low-frequency regime of $D(\omega) \sim \omega^{5}$. This shows a crossover as the size of the system is increased beyond a length scale, with the low-frequency VDoS showing a significant change in the power-law, with the exponent varying from $4.5$ to $5$ in two dimensions.

The enhanced stability in the VDoS at large system sizes suggests that quasilocalized modes in the bulk result in the power-law $\omega^5$, whereas modes localized on the surface are much softer and more unstable. We show further analysis comparing bulk and boundary localization in the next section. Finally, such behavior in the thermodynamic limit of the OBC systems under consideration is consistent with earlier studies in PBC systems~\cite{krishnan2022universal}.

\begin{figure}[t!]
\begin{center}
\includegraphics[width=0.48\textwidth]{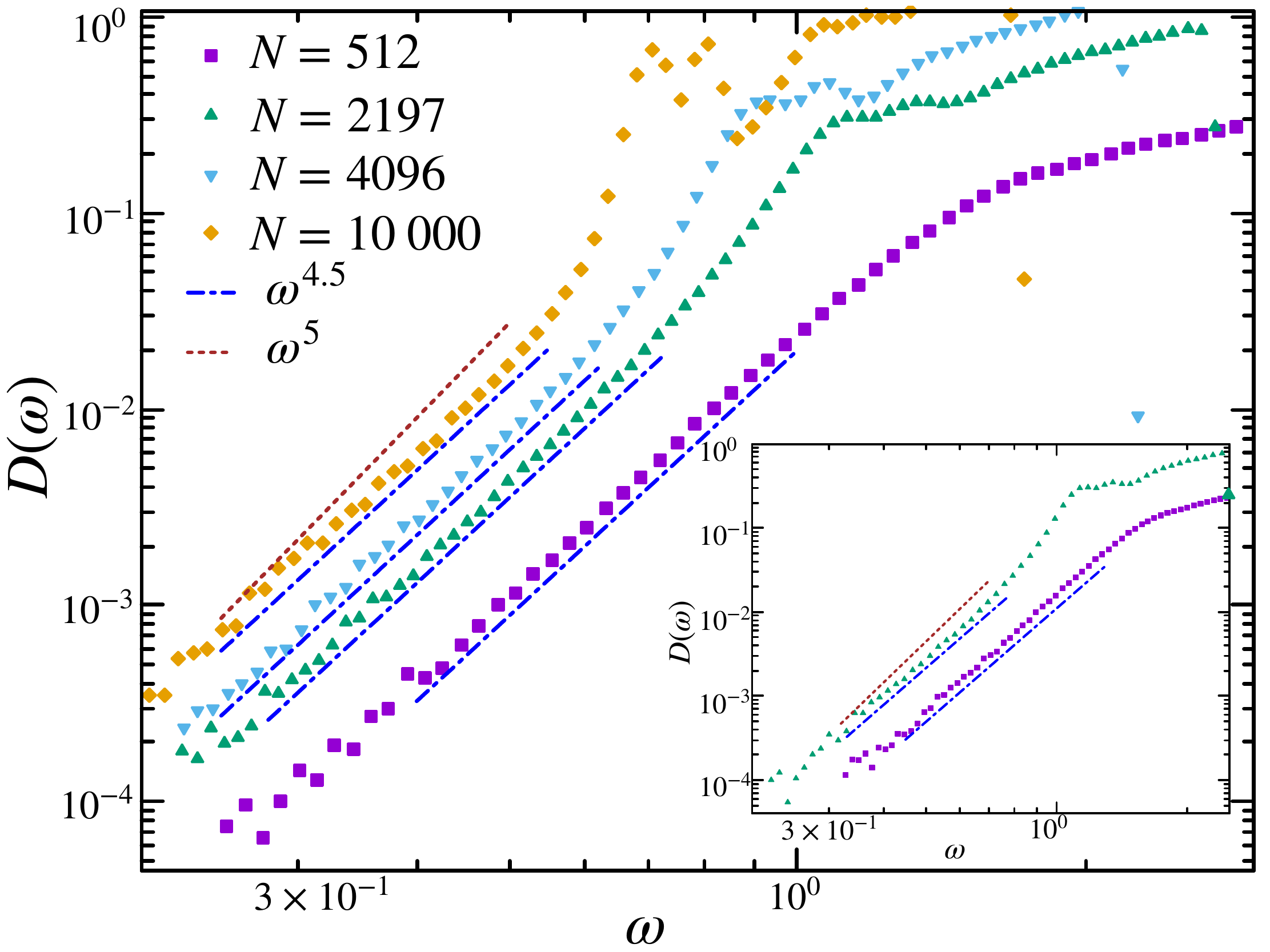}
\caption{System size dependence of the low-frequency vibrational spectrum of three-dimensional open boundary solids generated by annealing ($\gamma=0.1$). We sample the first $100$ non-zero eigenvalues of $2\times10^5$ configurations for system sizes $N = 512,~2197$ and $4096$. For the larger system of size $N=10,000$, the histogram is generated by sampling $8\times10^4$ configurations. \textbf{(Inset)} Distribution of low-frequency modes of various system sizes of three-dimensional solids generated via much slower annealing ($\gamma=0.01$). These distributions are drawn by sampling the first $100$ low-frequency modes of at least $80,000$ configurations for each system size. Solids corresponding to all the system sizes and degrees of annealing studied display a $D(\omega) \sim \omega^{4.5}$ in the low-frequency regime of their VDoS.}\label{fig_system_size_3d}
\end{center}
\end{figure}

In Fig.~\ref{fig_system_size_3d}, we plot the vibrational density of states for different system sizes in three dimensions. Here we observe a low-frequency behavior of $D(\omega) \sim \omega^{4.5}$ in contrast to the results in two dimensions. In the next section, we perform further analyses to examine this behavior. Moreover, as may be seen in the inset of Fig.~\ref{fig_system_size_3d}, we further confirm the robustness of this result by examining the VDoS of solids generated at a much higher degree of annealing by utilizing a damping constant that is one order of magnitude smaller ($\gamma = 0.01$).

\section{Eigenmode~Localization~---~Surface~vs~Bulk}\label{sec_bulk_sur}

In this section, we provide further analysis of the lowest frequency vibrational modes in two dimensions in order to determine the potential source of the stable vibrations achieved under open boundary conditions.

The spatial extent of a mode may be evaluated by measuring the participation ratio (PR), 
\begin{equation}
PR=\frac{(\sum_{i=1}^N\left\langle\psi_{i} \mid \psi_{i}\right\rangle)^2}{N\sum_{i=1}^N(\left\langle\psi_{i} \mid \psi_{i}\right\rangle)^2}\label{eqn_PR}
\end{equation} 
where $|\psi_{i}^{k}\rangle$ denotes the $d$ dimensional component of the eigenmode $\psi^k$, corresponding to particle $i$. System spanning modes possess a $PR$ close to unity, while spatially localized modes display a $PR \sim 1/N$.

\begin{figure}[t]
    \begin{center}
    \includegraphics[width=0.48\textwidth]{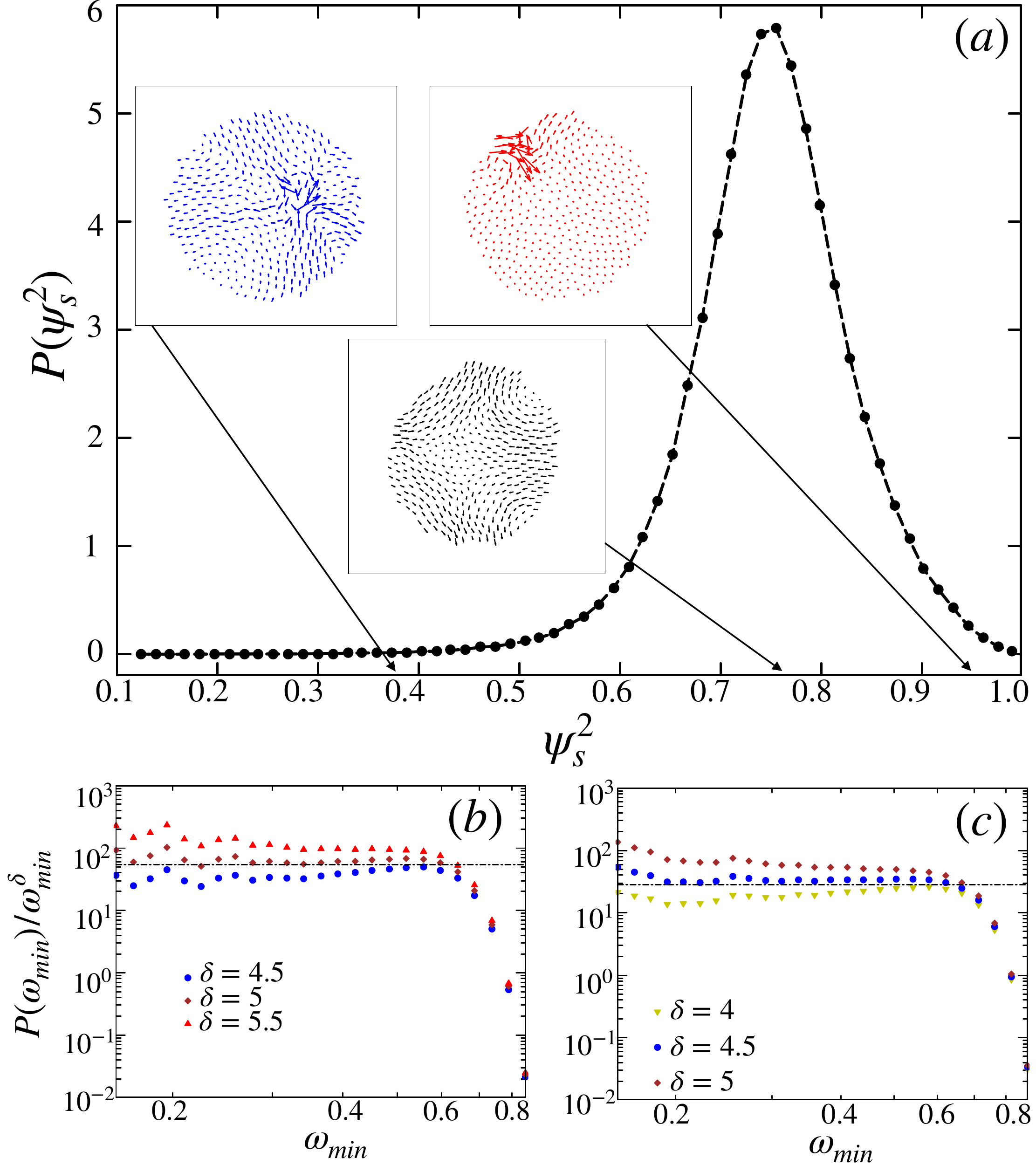}
    \caption{
    \textbf{(a)} Distribution of `Surface Participation' (SP) for the lowest frequency modes in two-dimensional solids of system size $N=400$. Modes localized on the surface of the solid display an $SP$ of $1$, whereas when $SP = 0$, the vibration is limited to the inner bulk. The peak of the distribution corresponds to extended modes spanning the whole solid. The distribution of the lowest frequency modes with predominantly \textbf{(b)} bulk and \textbf{(c)} surface localizations. Here, modes corresponding to an $SP \leq 0.7$ are considered to be bulk-localized vibrations and modes with an $SP \geq 0.8$ are considered surface-localized vibrations. Both histograms are constructed using $10^5$ lowest frequency modes. The bulk-localized modes display a $\delta=5$ whereas surface-localized modes show a $\delta=4.5$. } \label{fig_surface_analysis}
    \end{center}
\end{figure}

The contribution of the particles on the surface of the solid to a particular mode may then be estimated through the ``Surface Participation'' ($SP$) as defined below.
\begin{equation}
    \psi_{s}^{2}(k) = \sum_{i \in \texttt{surface}}\left\langle\psi_{i}^{k} \mid \psi_{i}^{k}\right\rangle
    \label{eqn_SP}
\end{equation}
The \texttt{surface} is defined as encompassing all the particles belonging to the boundary layer as determined by the stress analysis described in the previous section. Modes localized on the surface of the solid display an $SP$ close to $1$, whereas a vibration that is localized deeper within the bulk of the solid incurs an $SP$ closer to $0$. In Fig.~\ref{fig_surface_analysis}~\textbf{(a)} we plot the probability distribution of SP for the lowest frequency modes in two-dimensional solids of system size $N=400$. The peak of the distribution corresponds to modes with extended, system-spanning vibrations that all particles participate in. By observation, we consider modes with an $SP \leq 0.7$ to be predominantly bulk-localized vibrations and modes with an $SP \geq 0.8$ to be predominantly surface-localized. In Fig.~\ref{fig_surface_analysis}~\textbf{(b)} and~\textbf{(c)} we plot histograms of the lowest frequency divided by $\omega_{min}^{\delta}$ in order to extract the power-law. We find that bulk-localized modes (Fig.~\ref{fig_surface_analysis}~\textbf{(b)}) display the stabilized $\delta \approx 5$ while the surface-localized modes (Fig.~\ref{fig_surface_analysis}~\textbf{(c)}) present a more unstable $\delta \approx 4.5$. This points to the possibility that the low-stress environment of the bulk contributes stable vibrations while the large stresses at the surface result in modes of lower stability. Furthermore, such a separation potentially explains the large degree of softness observed in the vibrational spectrum at small system sizes.

\begin{figure}[t]
    \begin{center}
    \includegraphics[width=0.48\textwidth]{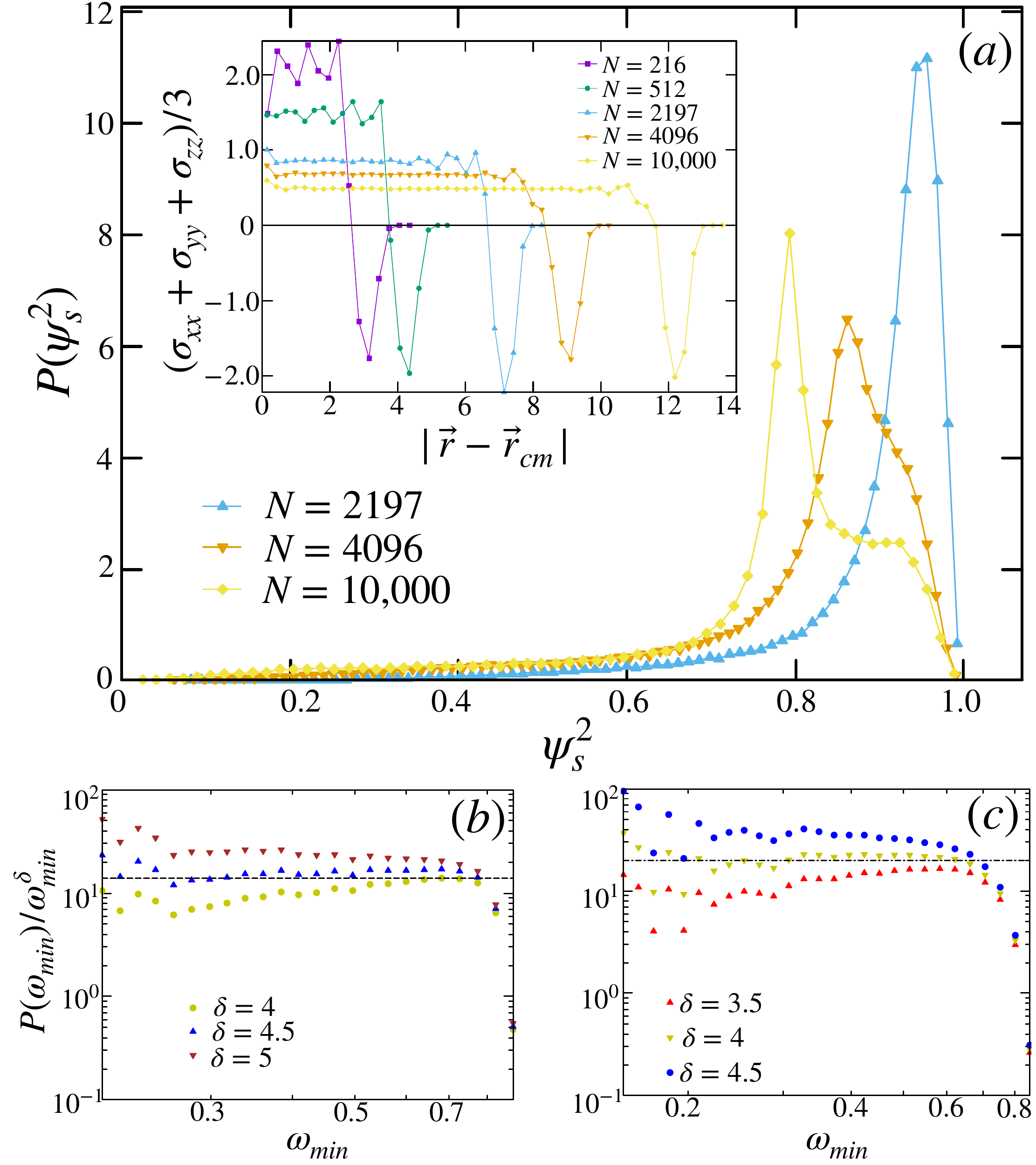}
    \caption{\textbf{(a)} Distribution of $SP$ for the lowest frequency modes in three-dimensional solids of system sizes $N = 2197, 4096$ and $10,000$. \textbf{(Inset)} Radial distribution of pressure for three-dimensional solids of various system sizes. In~\textbf{(b)} and~\textbf{(c)} we plot $P(\omega_{min})$ divided by $\omega_{min}^{\delta}$ for bulk and surface localized modes respectively. Modes corresponding to $SP\leq0.82$ are considered bulk-localized vibrations, and modes with $SP\geq0.9$ are considered to be surface-localized. Both histograms are constructed using samples of the lowest frequency modes from $5\times10^4$ configurations. Bulk-localized modes display a $\delta \approx 4.5$ whereas surface dominated modes show $\delta \approx 4$.}\label{fig_3d_sur}
    \end{center}
\end{figure}

We now provide a characterization of the lowest frequency vibrational modes in  three-dimensional open boundary solids. We study the contribution of surface particles on these modes through their ``Surface Participation'' ($SP$) as defined in~ Eq.~\eqref{eqn_SP}. In Fig.~\ref{fig_3d_sur}~\textbf{(a)}, we plot the probability distribution of SP corresponding to the lowest frequency modes in three-dimensional solids of various system sizes. The modes with a value of $SP$ corresponding to the peaks of the distributions are extended, system-spanning vibrations. In the case of the distribution corresponding to a system size $N = 4096$, we define the modes with $SP \leq 0.82$ as being predominantly bulk-localized vibrations and modes with $ \geq 0.9$ as predominantly surface-localized. In Fig.~\ref{fig_3d_sur}~\textbf{(b)} and~\textbf{(c)}, we plot histograms of the lowest frequency for bulk and surface localized modes, respectively. We find that bulk-localized modes display a $\delta \approx 4.5$ and surface-localized modes show a $\delta \approx 4$.

\section{Effect of Confining Stresses}\label{sec_confinement}

In this section, we show that the vibrational properties of systems under periodic boundaries can by reproduced by imposing macroscopic stresses on open boundary solids.

The stresses in a system are described through the force moment tensor, defined as
\begin{equation}
    \Sigma_{\alpha \beta}=\sum_{i} \sum_{j} f_{\alpha}^{i j} r_{\beta}^{i j}=\sigma_{\alpha \beta} A
\end{equation}
where $f_{\alpha}^{i j}$ is the $\alpha$-component of the force on particle $i$ by particle $j$, $r_{\beta}^{i j}$ is the $\beta$-component of the vector distance between the particles $i$ and $j$, $\sigma_{\alpha\beta}$ is the macroscopic stress tensor and $A$ is the area of the two dimensional system. Solids prepared under open boundary conditions have precisely zero macroscopic stresses, i.e. $\sigma_{xx}=\sigma_{yy}=\sigma_{xy}=0$. As described in the previous sections, such systems display a $D(\omega) \sim \omega^{\delta}$ with $\delta>4$, in contrast to PBC systems where $\delta = 4$. It is, therefore, interesting to analyze the effects of re-introducing stresses on the vibrational spectrum of solids.

\begin{figure}[t]
    \begin{center}
    \includegraphics[width=0.45\textwidth]{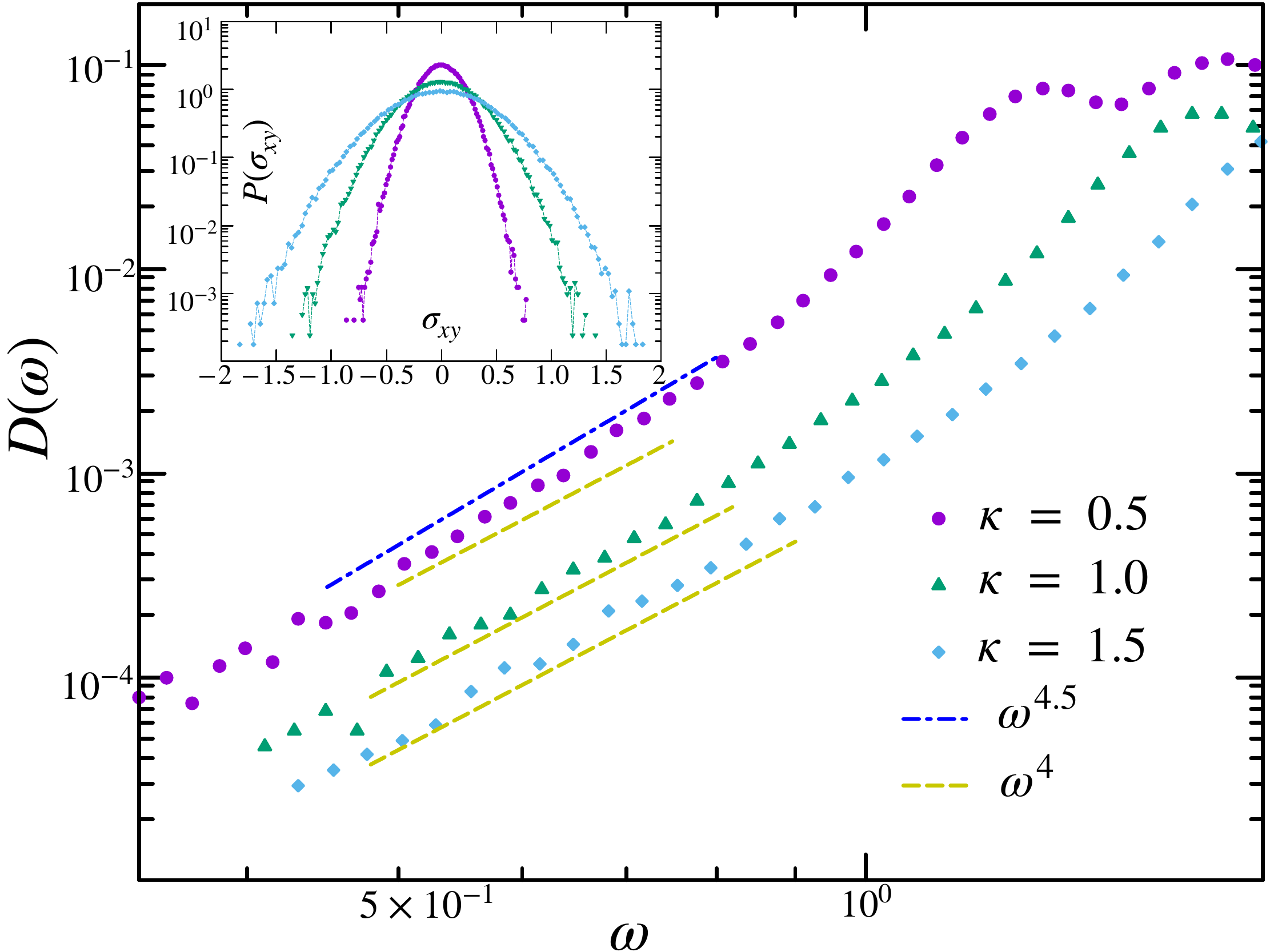}
    \caption{ 
    Effect of confinement on the vibrational spectrum of two-dimensional solids of system size $N=900$. We have sampled eigenvalues of $10^5$ configurations for different stiffness constants. (\textbf{Inset}) Shows the sample-to-sample distribution of shear stress of the solids for different stiffness of the harmonic interaction. Area $A$ of the confined solids is chosen from the radial stress distribution of the solids. Increasing the stiffness constant results in large shear stress fluctuation on the solid ensemble, which changes the VDoS from $D(\omega) \sim \omega^5$ to $D(\omega) \sim \omega^4$ in the low-frequency regime of the solids.
    }\label{fig_confinement_DoS}
    \end{center}
\end{figure}

In order to accomplish this, we have studied the vibrational modes of solids confined in a radially symmetric harmonic trap centered at the center of mass (CM) of the system, as follows:
\begin{footnotesize}
\begin{equation}
     U\left(\mathbf{r}_{1}, \mathbf{r}_{2}, \ldots \mathbf{r}_{n}\right) =  V\left(\mathbf{r}_{1}, \mathbf{r}_{2}, \ldots \mathbf{r}_{n}\right) + \frac{1}{2}\sum_{i=1}^{N} \kappa |\mathbf{r}_{i}-\mathbf{r}_{cm}|^2,
\end{equation}
\end{footnotesize}
where $\kappa$ is the spring constant of the trap, $\mathbf{r}_{cm}$ is the location of the CM and $V\left(\mathbf{r}_{1}, \mathbf{r}_{2}, \ldots \mathbf{r}_{n}\right)$ represents the pair-wise particle interaction. We then perform Damped Dynamics to obtain energy-minimized solid configurations.
Since the center of the harmonic trap is fixed, it breaks translational symmetry. However, it being radially symmetric, the system retains one zero mode corresponding to global rotations. The translational zero-modes are substituted with eigenvalues corresponding to the stiffness of the confining harmonic potential. We, therefore, sample the low-frequency modes while excluding the zero modes as well as this trivial addition to the spectrum. In Fig.~\ref{fig_confinement_DoS}, we display the effect of confinement on the low-frequency vibrational spectrum at various strengths of the harmonic trap. Remarkably we find that confinement results in a higher degree of softness in the vibrational spectrum of the solids. Increasing the stiffness constant results in large shear stresses in the solid ensemble with a corresponding change in the VDoS from $D(\omega) \sim \omega^5$ to $D(\omega) \sim \omega^4$.

\section{Conclusion and Discussion}\label{sec_conclusion}

In this paper, we have performed a detailed characterization of the vibrational modes of amorphous solids prepared under open boundary conditions in both two and three dimensions. We showed that structures prepared under open boundary conditions differ crucially in their stability properties in comparison to solids prepared with periodic boundary conditions. Specifically, we observed that the $D(\omega) \sim \omega^{\delta}$ with $\delta = 4$ seen in systems under PBC is modified to a $\delta \approx 5$ in 2D and $\delta \approx 4.5$ in 3D, for solids under OBC.
This points to the fact that open boundary solids which lack any macroscopic stresses are inherently {\it more stable} than their periodic boundary counterparts. These results reinforce the phenomenon observed by Krishnan et.~al.~\cite{krishnan2022universal} where shear-stabilized $(\sigma_{\alpha\beta} = 0 ~~ \forall ~~ \alpha \neq \beta)$ configurations under PBC display an increase in the exponent to $\delta \approx 5$. Further, we probed the nature of energy minima of systems under OBC through an analysis of the anharmonic coefficients associated with their lowest frequency vibrational mode. Surprisingly we have found that the anharmonic stability of the minimum is sensitive to the protocols employed in generating solid configurations. Moreover, the nature of the minima under consideration is also correlated with a change in the exponent of the VDoS. Specifically, the ensembles with unstable minima correspond to a $D(\omega) \sim \omega^4$, whereas ensembles with stable minima show larger exponents. In particular, we observed clear enhancements in the stability of configurations resulting from dissipative dynamical processes that \textit{anneal} the system when compared to those obtained via quenching minimization protocols. Next, we also characterized the dependence of the low-frequency modes on the system size. Solids at small systems sizes show a higher degree of softness in their low-frequency vibrations, as evidenced by a decrease in the exponent of the VDoS. We also performed an analysis of eigenmode localizations on the boundary and bulk of the system in order to decipher the source of the enhanced stabilization in systems under OBC. We found strong indications that the bulk of the solid predominantly contains stable vibrations, whereas the surface supports unstable modes. Finally, in order to isolate the cause of instability under PBC, we studied the vibrational spectrum of confined solids using a harmonic trap that introduces finite macroscopic stresses in the solid. We indeed showed that the $D(\omega) \sim \omega^4$ behavior of the VDoS is recovered with an increase in the strength of the confinement.
This is consistent with recent work where it has been shown that an increasing strain corresponds to an increase in the propensity for low-frequency modes~\cite{kriuchevskyi2022predicting}.

Our study highlights aspects of the protocols used to create amorphous structures that have direct consequences on the nature of the sampled energy minima. We showed that various energy-minimization procedures populate qualitatively different minima of the landscape, as has also been observed in previous studies~\cite{nishikawa2022relaxation, angelani2003general}. Our work in the context of open-boundary solids sheds light on the sparse nature of stable minima in such realistic systems. These results further raise important questions about the limits of stability that may be achieved through sufficient annealing.

Through our analysis of the stress profiles of solids under OBC, we found that the crossovers between the stable and unstable behaviors of the VDoS are well-described by a bulk-boundary decomposition of the solids. The enhanced stability is derived from the low-stress environment of the bulk of the solid supported by constant boundary stress. These results point to the fact that the \textit{frozen-in stresses} are crucial in determining the stability of the amorphous solids as well as their vibrational density of states. It would be interesting to use the characterization of the boundary layer and stress distributions that appear in such confined systems to understand the correlations between the macroscopic stress tensor and the distribution of eigenmodes.

Notably, our ensembles exhibit liquid-like surface tensions, as seen by the conformance of bulk pressures to the Laplace law. This observation bares further examination in order to probe any correlation to the enhanced stability of open-boundary solids. Although our results for the cut-out protocol reveal striking properties of vibrations in solids under OBC, it would be intriguing to further characterize this behavior with other preparation protocols, such as by evaporation and deposition, which is of direct experimental relevance. Finally, our results in three-dimensional systems regarding their smaller power-law exponent $D(\omega) \sim \omega^{4.5}$ in comparison to two-dimensional systems is surprising in the context of earlier work displaying a universal $\omega^{5}$ behavior in shear-stabilized systems under PBC. This points to the possibility that systems in higher dimensions support more localization of vibrations. It would be interesting to explore aspects of this non-universality in open-boundary systems.

\begin{acknowledgments}
We thank Roshan Maharana, Monoj Adhikari, Deepak Dhar and Prasad Perlekar for useful discussions. S. K. would like to acknowledge support through the Swarna Jayanti Fellowship Grant No. DST/SJF/PSA-01/2018-19 and SB/SFJ/2019-20/05. The work of K.~R. was partially supported by the SERB-MATRICS grant MTR/2022/000966. This project was funded by intramural funds at TIFR Hyderabad under Project Identification No. RTI 4007 from the Department of Atomic Energy, Government of India.
\end{acknowledgments}

\bibliographystyle{apsrev4-2} 
\bibliography{VDos_open_Bibliography}
\end{document}